\documentclass[floatfix,aps,twocolumn,showpacs,nofootinbib,superscriptaddress]{revtex4}

\usepackage{graphicx}
\usepackage{epsfig}

\usepackage{amsmath}
\usepackage{amsfonts}
\usepackage{amssymb}
\usepackage{bm}
\usepackage{mathtools}

\usepackage{hyperref}
\usepackage{dsfont}
\usepackage{lipsum}

\usepackage{tikz}
\usepackage{circuitikz}
\usetikzlibrary{arrows, shapes}
\usepackage{pgfplots}
\pgfplotsset{compat=1.14}

\usepackage[final]{pdfpages}

\newcommand{\figpath}[1]{./figures/#1}

\newcommand{\mean}[1]{\left\langle #1 \right\rangle}

\newcommand{\ket}[1]{| #1 \rangle}
\newcommand{\bra}[1]{\langle #1 |}

\newcommand{\Li}{\operatorname{Li}}

\begin{document}
\author{Nahuel Freitas}
\affiliation{Complex Systems and Statistical Mechanics, Department of Physics and Materials Science,
University of Luxembourg, L-1511 Luxembourg, Luxembourg}
\author{Karel Proesmans}
\affiliation{Complex Systems and Statistical Mechanics, Department of Physics and Materials Science,
University of Luxembourg, L-1511 Luxembourg, Luxembourg}
\affiliation{Hasselt University, B-3590 Diepenbeek, Belgium}
\author{Massimiliano Esposito}
\affiliation{Complex Systems and Statistical Mechanics, Department of Physics and Materials Science,
University of Luxembourg, L-1511 Luxembourg, Luxembourg}

\title{Reliability and entropy production in non-equilibrium electronic memories}

\date{\today}

\begin{abstract}
We find the relation between reliability and entropy production
in a realistic model of electronic memory (low-power MOS-based SRAM)
where logical values are encoded as metastable non-equilibrium states.
We employ large deviations techniques to obtain an analytical expression
for the bistable quasipotential describing the non-equilibrium steady
state and use it to derive an explicit expression bounding the error
rate of the memory. Our results go beyond the dominant contribution
given by classical instanton theory and provide accurate estimates
of the error rate as confirmed by comparison with stochastic
simulations. The methods developed can be adapted to study the
reliability of broad classes of nonlinear devices subjected to thermal
noise.
\end{abstract}

\maketitle

\section{Introduction}

A common strategy to reduce the energy consumption of electronic computing devices
is to reduce the voltage at which they are powered. However, this strategy is limited
by the fact that as the operation voltage is reduced, different sources of
electrical noise start to play an increasingly important role
\cite{natori1998, wang2006, krishnan2007, rezaei2020}. The most fundamental
and unavoidable one is given by the thermal fluctuations intrinsic to
any device. It originates from the interaction with degrees of freedom that are not explicitly described, but that can be normally assumed to be at thermal equilibrium.
A rigorous description of intrinsic thermal noise
in complex and non-linear electronic circuits is thus a fundamental problem in
modern engineering, of great importance for the search of new efficient computing schemes \cite{han2013,krishnan2007, rezaei2020, gu2019, gu2020}.
However, it is also a hard problem that is usually given approximate
treatments involving different kinds of approximations that are difficult to control, and that in general compromise thermodynamic consistency \cite{hanggi1988, freitas2020}.
This issue was recently addressed by the development of a general theoretical framework to construct thermodynamically consistent stochastic models of non-linear electronic circuits \cite{freitas2020}.

In this work we make use of that framework to analyze the tradeoff
between reliability and dissipation (i.e. entropy production)
of low-power static random access memory (SRAM) cells.
Due to their speed and low energy consumption, SRAM cells are employed
as internal memory in virtually all modern processors.
The occurrence of errors induced by thermal noise in low-power implementations has been mainly studied using numerical methods based on stochastic simulations \cite{li2006, rezaei2020}. The reason is that in low-power regimes current fluctuations are Poissonian and cannot be faithfully described as Gaussian noise \cite{sarpeshkar1993}, which considerably complicates analytical treatments.
However, since one is typically interested in determining the rate of errors
in regimes where errors are rare, the amount of computational time demanded by the
stochastic simulations can be extremely large \cite{rezaei2020}.
In this contribution we report two main results. First, we obtain an analytical description of the steady state fluctuations of the memory, fully capturing the non-equilibrium transition from a monostable phase
into the bistable phase that allows the representation of a bit.
Secondly, we show how to employ the previous result to analytically estimate the error rate of the memory. By comparing with exact stochastic simulations, we show that our analytical estimation correctly describes the scaling of the error rate with the voltage that powers the memory. Then, we show that the error rate is
exponentially suppressed as the square of the dissipation (for large dissipation). To get there, we make use of advanced methods
from stochastic thermodynamics \cite{freitas2020}, large deviations theory \cite{elgart2004,kamenev2008, touchette2009, assaf2017, limmer2021}, and first-passage time statistics \cite{hanggi1990, redner2001, van1992}.

\section{Basic Model}

\begin{figure}
\centering
\ctikzset{bipoles/length=1.1cm}
\begin{circuitikz}[scale=.7, thick]

\begin{scope}[xshift=-6.5cm]
\draw (0,0) node[circ, label=north:$\bar b$]{} -- ++(.6,0) -- ++(0,.6) -- +(.15,0) node[american not port, anchor=in](not1){};
\draw (not1.out) -- ++(.15,0) -- ++(0,-.6) -- ++(.6,0) node[circ, label=north:$b$]{};
\draw (.6,0) -- ++(0,-.6) -- +(.15,0) node[american not port, xscale=-1, anchor=out](not2){};
\draw (not2.in) -- ++(.15,0) -- ++(0,.6);
\end{scope}

\draw (0,1) node[pigfete] (pfet1) {};
\draw (0,-1) node[nigfete] (nfet1) {};
\draw (pfet1.S) -- ++(0,.2) node[vcc]{$V_\text{dd}$};
\draw (nfet1.S) -- ++(0,-.2) node[vee]{$-V_\text{dd}$};
\draw (nfet1.G) -| (-1.5,0);
\draw (pfet1.G) -| (-1.5,0);
\draw (-2,0) -- (-2.5,0) node[circ, label=south:$v_2$]{};
\draw (-1.5,0) -- ++(-.5,0);
\draw (nfet1.D) -- (0,0);
\draw (pfet1.D) |- (.5,0);
\begin{scope}[xshift=2cm, xscale=-1]
\draw (0,1) node[pigfete, xscale=-1] (pfet2) {};
\draw (0,-1) node[nigfete, xscale=-1] (nfet2) {};
\draw (pfet2.S) -- ++(0,.2) node[vcc]{$V_\text{dd}$};
\draw (nfet2.S) -- ++(0,-.2) node[vee]{$-V_\text{dd}$};
\draw (nfet2.G) -| (-1.5,0);
\draw (pfet2.G) -| (-1.5,0);
\draw (-1.5,0) -- ++(-.5,0);
\draw (nfet2.D) |- (.5,0);
\draw (pfet2.D) |- (.5,0);
\end{scope}
\draw (-2,0) -- ++(0,-1.8) -- ++(1.5,0) to[xing] ++(1,0) -- ++(1,0) -- ++(0,1.8);
\draw (.5,0) -- ++(0,1.8) -- ++(1,0) to[xing] ++(1,0) -- ++(1.5,0) -- ++(0,-1.8) -- ++(.5,0) node[circ, label=south:$v_1$]{};

\begin{scope}[yshift=-6.5cm, xshift=-5cm]
\draw (-1,9.0) node {(a)};
\draw (-1,2.2) node {(b)};
\draw (0,0) node[nigfete, anchor=center] (nfet) {};
\draw (nfet.G) node[circ, label=west:G] {};
\draw (nfet.S) node[circ, label=south:S] {};
\draw (nfet.D) ++(0,-.3) to[short, i_<=$I_D$] ++(0,.8) node[circ, label=north:D] {};
\draw (nfet.B) node[circ, label=east:B] {};

\draw[-latex, line width=3] (1.6,0) -- (2.6,0);

\begin{scope}[xshift=4.5cm]
\draw (0,-1) to[C=$C_g$] (0,1) to[short] ++(-.5,0) node[circ, label=west:G]{};
\draw (2.2,-1) to[european resistor] (2.2,1) to[short, i_<=$I_D$] ++(0, .7) node[circ, label=north:D]{};
\draw[latex-] (1.75,-.5) -- (1.75,.5) node[midway, left]{$\lambda^n_+$};
\draw[-latex] (2.65,-.5) -- (2.65,.5) node[midway, right]{$\lambda^n_-$};
\draw (4.2,-1) to[capacitor, l_=$C_o$] (4.2,1);
\draw (4.2,-1) to[short] (2.2,-1);
\draw (4.2,1) to[short] (2.2,1);
\draw(0,-1) to[short] (2.2,-1);
\draw(1.1,-1) to[short] ++(0,-.5) node[circ, label=west:S, label=east:B]{};
\draw (2.2,0) node[]{$n$};
\end{scope}
\end{scope}

\end{circuitikz}
\caption{(a) A bistable logical circuit constructed with two NOT gates, representing a bit, and its CMOS implementation, where each NOT gate is constructed with one $p$MOS (top) and one $n$MOS (bottom)
transistors. (b) Each transistor (in this example an $n$MOS one) is modelled as a conduction
channel between drain (D) and source (S) terminals, with associated rates $\lambda_\pm^n$. The gate-body (G-B) interface is represented as a capacitor $C_g$, and another capacitor $C_o$ takes into account the output capacitance. Other parasitic capacitances could also be taken into account,
for example between drain and gate. With this model and taking $V_\text{ss} = -V_\text{dd}$, the total electrostatic energy of the full circuit
is $\Phi(v_1, v_2) = (C/2)(v_1^2 + v_2^2) + C V_\text{dd}^2$, with $C = 2(C_o + C_g)$.}
\label{fig:model}
\end{figure}
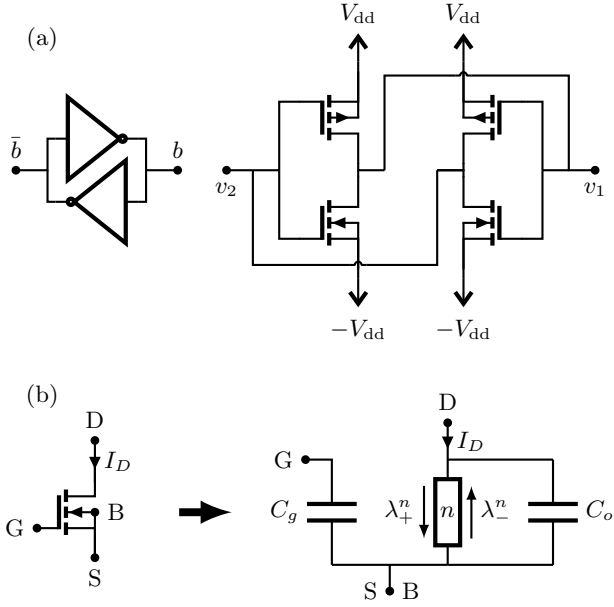

We consider the usual model of a SRAM memory cell core: two inverters, or NOT gates, connected in a loop (see Figure \ref{fig:model}-(a)). In particular, we consider the implementation based on complementary metal-oxide-semiconductor (MOS) transistors. In this case, each inverter is itself composed of an $n$MOS transistor and a $p$MOS transistor. The circuit is powered by applying a voltage bias $\Delta V = V_\text{dd} - V_\text{ss}$. The deterministic and linear stability analysis of the circuit (see Appendix \ref{ap:deterministic}) shows that
for low values of $\Delta V$ the circuit has a unique steady state, but when $\Delta V$ is
above a critical value there is a transition into bistability, which is employed to encode
a single bit of information.
The transistors are modelled as externally controlled conduction channels with associated
capacitances (see Figure \ref{fig:model}-(b)). The charge conduction through each transistor channel is modelled as a
bidirectional Poisson process.
Thus, to each ($n$/$p$)MOS transistor we associate two
Poisson rates $\lambda_\pm^{n/p}(V_\text{GS}, V_\text{DS})$, where the subindices
$\pm$ correspond to the forward and backward conduction directions, and $V_\text{GS}$
and $V_\text{DS}$ are the gate-source and drain-source voltage drops, respectively.
For fixed voltages $V_\text{dd}$ and $V_\text{ss}$ the circuit has two independent
degrees of freedom: the voltages $v_1$ and $v_2$ at the outputs of each inverter.
These are discrete stochastic quantities, that in principle can only take the
values $m v_e$, where $m$ is any integer and $v_e = q_e/C$ ($q_e$ is the positive electron
charge and $C$ a value of capacitance characterizing the device, see Figure \ref{fig:model}-(b)).

At any given time the state of the system is described by a probability distribution
$P(v_1, v_2, t)$ over the state space. Its evolution is given by the following master
equation
\begin{equation}
\begin{split}
d_t P(v_1, v_2, t)
&= PA|_{v_1-v_e, v_2} + PB|_{v_1+v_e, v_2} \\
&+ PA^*|_{v_1, v_2-v_e} + PB^*|_{v_1, v_2+v_e} \\
&- P (A + B + A^* + B^*)|_{v_1, v_2},
\end{split}
\label{eq:master}
\end{equation}
where we are using the compact notation $PA|_{v_1,v_2}= P(v_1,v_2)A(v_1,v_2)$,
and $A^*(v_1,v_2)=A(v_2,v_1)$.
The transition rates $A(v_1, v_2)$ and $B(v_1,v_2)$ are combinations of the Poisson
rates assigned to the transistors:
\begin{equation}
\begin{split}
A(v_1,v_2) &= \lambda_+^p(v_1,v_2) + \lambda_-^n(v_1,v_2) \\
B(v_1,v_2) &= \lambda_-^p(v_1,v_2) + \lambda_+^n(v_1,v_2).
\label{eq:rates_up_down}
\end{split}
\end{equation}
In order to guarantee thermodynamic consistency, the  Poisson rates $\lambda_\pm^{n/p}(v_1,v_2)$
must satisfy the so called \emph{local detailed balance (LDB)} conditions. As an example,
for the $p$MOS transistor in the first inverter, this condition reads:
\begin{equation}
\frac{\lambda_+^{p}(v_1, v_2)}{\lambda_-^{p}(v_1+v_e, v_2)} =
e^{-\delta Q/(k_bT)},
\label{eq:ldb}
\end{equation}
where $\qquad \delta Q = \Phi(v_1+v_e, v_2)- \Phi(v_1,v_2) - q_e V_\text{dd}$,
$\Phi(v_1, v_2)$ is the electrostatic energy of the system,
and we have considered the environment of the transistor to be in equilibrium at temperature $T$.
For $V_\text{ss} =  - V_\text{dd}$,
as we will consider in the following, the electrostatic energy
is $\Phi(v_1, v_2) = (C/2)(v_1^2 + v_2^2) + C V_\text{dd}^2$. Thus the LDB condition of Eq. \eqref{eq:master} relates the rates of the transitions $v_1 \rightleftarrows v_1 + v_e$
to the difference in internal energy between those states, and the work $q_e V_\text{dd}$
realized by the voltage sources during the forward transition. Then,
$\delta Q$ is the total energy change associated to that transition, and since by energy
conservation it must be provided by the environment of the device, it is the heat interchanged with it. A condition analogous to Eq. \eqref{eq:ldb} is imposed
to all the transistors present in the circuit. A general procedure to construct
thermodynamically consistent rates based on the I-V curve characterization of a given
devices was recently identified in \cite{freitas2020}. For the case of MOS transistors
in subthreshold operation, one obtains:
\begin{equation}
\begin{split}
\lambda_+^p(v_1,v_2) &= (I_0/q_e) \: e^{(V_\text{dd}-v_2-V_\text{th})/(\text{n}V_T)} \qquad\\
\lambda_-^p(v_1,v_2) &= \lambda_+^p(v_1,v_2) \: e^{-(V_\text{dd}-v_1)/V_T} \: e^{-(v_e/2)/V_T},
\end{split}
\label{eq:mos_rates}
\end{equation}
and $\lambda_{\pm}^n(v_1,v_2) = \lambda_{\pm}^p(-v_1,-v_2)$.
In the previous equation $V_T = k_bT/q_e$ is the thermal voltage
and $I_0$, $V_\text{th}$, and $\text{n}$
are parameters characterizing the transistor
(respectively known as \emph{specific current},
\emph{threshold voltage}, and \emph{slope factor}).
An incorrect procedure to construct transition rates, which is however used
in some numerical simulations \cite{li2006, rezaei2020},
is to employ the rates directly obtained from the I-V curve characterization,
without enforcing the LDB conditions.
In that way one finds rates that are obtained from the ones of Eq. (\ref{eq:mos_rates})
by removing the factor $e^{-(v_e/2)/V_T}$ appearing in $\lambda_-^{p/n}$. Although this
factor is in many situations very close to $1$, it can become relevant for small
devices or at low temperatures, and it is in fact responsible for the charging effects
in single-electron devices \cite{fulton1987, pothier1992, devoret1990}.
Also, neglecting that factor leads to systematic errors in the determination of the steady state.
For example, in modern CMOS fabrication processes capacitance values
as low as $C\simeq 50$ aF can be attained \cite{zheng2016}, which correspond to elementary voltages as high as $v_e \simeq 3$ mV. At room temperature we have $V_T \simeq 26$ mV and therefore $v_e/V_T \simeq 0.1$ and $e^{-(v_e/2)/V_T}\simeq 0.95$.

For mathematical simplicity, the parameters $I_0$, $V_\text{th}$ and $\text{n}$ are considered
to be the same for all the four transistors involved in the circuit. That is,
we are not taking into account any variability associated with the fabrication process \cite{mukhopadhyay2004}. It should be possible to extend our results to systems with asymmetric parameters.


\section{Steady state distribution and Large Deviations Principle}

To find the steady state of the memory one option is to
construct the generator of the master equation in Eq. \eqref{eq:master} and compute its eigenvector of zero eigenvalue. Analytical progress is possible
by considering a macroscopic limit and employing the principle of large deviations.
This limit consists in assuming that the elementary voltage $v_e$ is negligible compared
to all other voltage scales, which in this case
are the thermal voltage $V_T$ and
the powering voltage $V_\text{dd}$ (thus, the limit $v_e \to 0$
used in the following must be interpreted as $v_e/V_T \to 0$
and $v_e/V_\text{dd} \to 0$ for fixed $V_T$ and $V_\text{dd}$).
Physically, this corresponds to large devices, for which the
typical capacitance $C$ is large and thus $v_e$ is small. Also, from Eq. \eqref{eq:mos_rates}
we have that the Poisson rates are proportional to $(I_0/C)v_e^{-1}$.
As explained in Appendix \ref{ap:large_dev},
the specific current $I_0$ can also be
considered to be proportional to the size of the device, and therefore
we see that the transition rates scale as $v_e^{-1}$. Under these conditions, as $v_e \to 0$,
the deterministic equations of motion are recovered from
the master equation in Eq. \eqref{eq:master} (see Appendixes \ref{ap:deterministic} and \ref{ap:large_dev}),
and one also expects
the distribution $P_\text{ss}(v_1,v_2)$ to become strongly peaked around the deterministic
stationary values \cite{van1992,cossetto2020}. In this context, the LD principle states that departures
from the deterministic values are suppressed exponentially in $v_e^{-1}$. This is expressed
mathematically as the existence of the limit
$f(v_1, v_2) = \lim_{v_e \to 0}\: -v_e \log(P_\text{ss}(v_1,v_2))$, or equivalently:
\begin{equation}
P_\text{ss}(v_1,v_2) \asymp e^{-(f(v_1,v_2) + o(v_e))/v_e}.
\label{eq:lda}
\end{equation}
Therefore, as $v_e \to 0$, the values of $v_1$ and $v_2$ will be perfectly localized
at a global minimum of the \emph{rate function} $f(v_1,v_2)$. Indeed, the minima of $f(v_1,v_2)$ correspond to the deterministic fixed points (see Appendixes \ref{ap:deterministic} and \ref{ap:large_dev}).
We will refer to the function $f(v_1,v_2)$ as a
\emph{quasipotential} describing
the steady state distribution. This is in analogy to an equilibrium situation, where the
steady state must be the equilibrium Boltzmann distribution
$P_\text{eq}(v_1,v_2) \propto \exp(-\Phi(v_1, v_2)/k_bT)$
and thus, by Eq. \eqref{eq:lda},
$f(v_1,v_2)$ should match the true potential energy $\Phi(v_1,v_2)$
scaled by the thermal voltage $V_T$. Also, the interpretation of $f(v_1, v_2)$ as a potential
has a deeper justification on the fact that it always is a Lyapunov function for the deterministic
dynamics \cite{cossetto2020}, as the true potential energy is for equilibrium settings.

\begin{figure*}
\begin{tikzpicture}
\node () at (0,0) {\includegraphics[width=\textwidth]{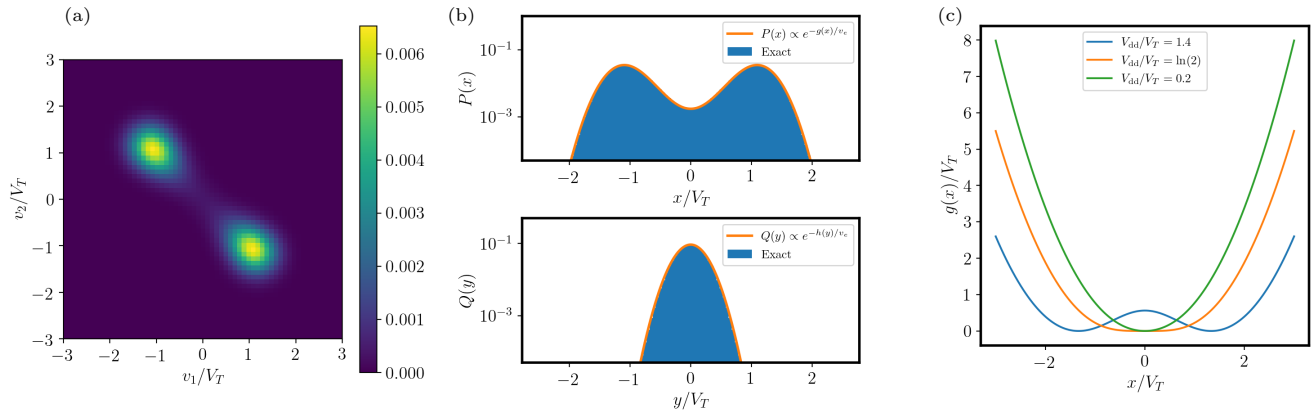}};
\node () at (-8,2.4) {\scriptsize (a)};
\node () at (-2.9,2.4) {\scriptsize (b)};
\node () at (3.6,2.4) {\scriptsize (c)};
\end{tikzpicture}
\caption{(a) Exact steady state obtained by numerical integration of the master equation
($V_\text{dd}/V_T=1.2$, $v_e/V_T=0.1$, $\text{n=1}$). (b) Partial distributions
for the variables $x$ and $y$ as obtained from the exact global distribution in (a), and
from the analytical results of Eq. \eqref{eq:dg_dh}. (c) Quasipotential $g(x)/v_e$ for different values of $V_\text{dd}$ ($v_e/V_T=0.1$, $\text{n=1}$).}
\label{fig:ss}
\end{figure*}

Plugging Eq. \eqref{eq:lda} into Eq. \eqref{eq:master},
imposing $d_t P_\text{ss} = 0$, and only keeping the lower order terms in $v_e$, we obtain
the following differential equation for $f(v_1,v_2)$:
\begin{equation}
\begin{split}
    0 &= \left(e^{\partial_{v_1} f}-1\right)  a(v_1, v_2) + \left(e^{-\partial_{v_1} f}-1\right)  b(v_1, v_2)  \\
    &+ \left(e^{\partial_{v_2}  f}-1\right) a(v_2, v_1) + \left(e^{-\partial_{v_2}  f}-1\right)  b(v_2, v_1)
     \label{eq:diff_eq}
\end{split}
\end{equation}

where $a(v_1,v_2) = \lim_{v_e \to 0} v_e A(v_1, v_2)$, and the same for $b(v_1, v_2)$.
Interestingly, the same equation can be obtained by more general path integral methods, in terms of a
Hamiltonian defining an action in the space of all possible stochastic trajectories \cite{kamenev2011, cossetto2020}.
This equation cannot be solved exactly. However, an approximate solution can be found
by exploiting the fact that the variables ${x=(v_1-v_2)/2}$ and ${y=(v_1+v_2)/2}$ are, appart of some trivial correlations discussed below, approximately independent.
Thus, as explained in Appendix \ref{ap:large_dev}, from Eq. \eqref{eq:diff_eq}
the rate functions $g(x)$
and $h(y)$ corresponding to the partial distributions
${P(x) = \sum_y P_\text{ss}(y+x,y-x) \asymp \exp(-g(x)/v_e)}$
and
${Q(y) = \sum_x P_\text{ss}(y+x,y-x) \asymp \exp(-h(y)/v_e)}$, can be found to
be
\begin{equation}
\begin{split}
  d_x g(x) &= 2\log\left(\frac{a(-x,0) + b(x,0)}{a(x,0) + b(-x,0)}\right) \\
  d_y h(y) &= 2\log\left(\frac{b(x_\text{min},y)+b(-x_\text{min},y)}
{a(x_\text{min},y)+a(-x_\text{min},y)}\right),
  \label{eq:dg_dh}
\end{split}
\end{equation}
where the change of variables in the functions $a$ and $b$ is understood, and
$x_\text{min}$ in the expression for $d_y h$ is the minimum of $g(x)$.
The variables $x$ and $y$ will be always correlated
because, since $v_1/v_e$ and $v_2/v_e$ are integer random variables, their difference $2x/v_e$
and sum $2y/v_e$ will always have the same parity. If, however, when restricted
to a given parity, $x$ and $y$ are independent, and if both parities have the same
probability, then the full probability
distribution $P_\text{ss}(v_1, v_2)$ can be reconstructed from the
partial distributions $P(x)$ and $Q(y)$ as
\begin{equation}
P_\text{ss}(y+x, y-x) =  2 P(x) Q(y) \text{Par}(x,y) ,
\label{eq:ss_parity}
\end{equation}
where $\text{Par}(x, y)$ is one if $2x/v_e$ and $2y/v_e$ have the same parity, or
zero if they do not.

The results in Eq. \eqref{eq:dg_dh} are in principle valid for any Poisson rates $\lambda_\pm^{n/p}$.
Remarkably, for the particular MOS rates of Eq. \eqref{eq:mos_rates}, the expression for $d_xg$ can be integrated exactly, resulting in:
\begin{equation}
g(x) = \frac{x^2\!+\!2V_\text{dd}\:x }{V_T}
+ \frac{2\text{n} V_T}{\text{n}+2} \left[ L(x, V_\text{dd}) \!-\!L(x, -V_\text{dd})\right],
\label{eq:exact_g}
\end{equation}
where $L(x,V_\text{dd}) = \Li_2\left(-\exp((V_\text{dd} + x(1+2/\text{n}))/V_T)\right)$, and $\Li_2(\cdot)$ is the polylogarithm function of second order. This is the
first important result of this work, and will allow us to analytically estimate
the error rate of a low-power SRAM memory cell in the next section.
In turn, the rate function $h(y)$ can be seen to satisfy $h(y) = h_0 \: y^2/V_T + \mathcal{O}(y^4)$
(an expression for $h_0$ in terms of the circuit parameters is given in Appendix \ref{ap:large_dev}).

In Figure \ref{fig:ss}-(a) we show the exact steady state distribution $P_\text{ss}(v_1, v_2)$
obtained by numerically evolving Eq. \eqref{eq:master} for $v_e/V_T = 0.1$, $V_\text{dd}/V_T = 1.2$, and
$\text{n}=1$.
We see that for these parameters the most probable values are distributed around
$v_1 = -v_2 \simeq \pm V_\text{dd}$, i.e., the possible solutions to the deterministic
equations of motion (Appendix \ref{ap:deterministic}). In Figure \ref{fig:ss}-(b)
we compare the exact partial distributions $P(x)$ and $Q(y)$ for the variables $x=(v_1-v_2)/2$ and
$y=(v_1+v_2)/2$, respectively, with the ones obtained from the quasipotentials
$g(x)/v_e$ and $h(y)/v_e$. We see that the agreement is remarkable despite the
value of $v_e$ being only one order of magnitude lower than $V_T$ and $V_\text{dd}$.
Finally, in Figure \ref{fig:ss}-(c) we show the quasipotential $g(x)$ for
different values of the powering voltage $V_\text{dd}$. We see that there is a
transition between a unimodal steady state and the bimodal distribution compatible
with bistability, that for $\text{n}=1$
happens at $V_\text{dd} = \ln(2) V_T$ (the data-retention voltage), as can be seen from the analysis of the
deterministic equations (Appendix \ref{ap:deterministic}).

\section{Error rate}

If the initial state of the system is close to one of the two possible metastable
NESSs, let us say $v_1 = -v_2 \simeq V_\text{dd}$, the ensuing dynamics will be characterized by two different time scales. First, a fast relaxation on the local
basin of attraction will take place. Indeed, from the deterministic equations (Appendix  \ref{ap:deterministic}) we
see that this relaxation develops at a rate
$\lambda_\text{eq} \simeq \tau_0^{-1} (v_e/V_T) \: e^{2V_\text{dd}/V_T}$ that increases exponentially
with $V_\text{dd}$, where $\tau_0 = (q_e/I_0) \: e^{V_\text{th}/(\text{n}V_T)}$ is a natural time scale for this problem.
After this local metastable NESS has been reached, a slow dynamics consisting of rare transitions to the other possible metastable NESS follows.
Since the metastable NESSs are associated to the values of the stored bit,
this rare transitions are considered errors. We are interested in computing the error
rate $\lambda_\text{err}$ in terms of the circuit parameters.
This is a hard problem that has been mainly
treated numerically \cite{li2006, rezaei2020}, and for which a rigorous stochastic treatment
is crucial. It is possible to see that, to leading order in $v_e^{-1}$, the rate of
escape out of a NESSs centered around $\mathbf{v}_\text{min} = (v_1^\text{min}, v_2^\text{min})$
can be obtained from the quasipotential $f$ thanks to the following result \cite{bouchet2016, cossetto2020}:
\begin{equation}
\lim_{v_e \to 0} \: v_e \log(\tau_0 \lambda_\text{err}) =
- (f(\mathbf{v}^*)- f(\mathbf{v}_\text{min})),
\label{eq:ld_rate}
\end{equation}
where $\mathbf{v}^*$ is a saddle point of the quasipotential (which in this case is $\mathbf{v}^*=(0,0)$).
The factor $\exp(-(f(\mathbf{v}^*)- f(\mathbf{v}_\text{min}))/v_e)$ is also the dominant
contribution to the probability of a trajectory, or `instanton', going from $\mathbf{v}_\text{min}$ to
$\mathbf{v}^*$ \cite{cossetto2020}. This result can be considered
a generalization to NESSs of the classical Arrhenius's law \cite{arrhenius1889},
and in this case leads to the `dominant' estimate of the error rate
\begin{equation}
\lambda_\text{err}^\text{D} = \tau_0^{-1} e^{-(g(0) - g(x_\text{min}))/v_e},
\label{eq:lerr_ld}
\end{equation}
that can be readily evaluated from Eq. \eqref{eq:exact_g}. However, this
estimate misses any contribution to $\lambda_\text{err}$ that is subexponential
in $v_e^{-1}$, but that might be anyway relevant for finite values of $v_e$.
For equilibrium systems some subexponential factors are provided by the
classic Eyring-Kramers formula \cite{eyring1935,kramers1940,berglund2011}, in terms of
the curvature of the energy surface at the fixed and saddle points.
For out of equilibrium systems with Gaussian noise, subexponential corrections
are discussed in \cite{bouchet2016, falasco2021}. In our case, since we are dealing with a
discrete out of equilibrium system subjected to shot noise, we resort to the general method
explained in the following.

\begin{figure*}[ht!]
\begin{tikzpicture}
\node () at (0,0) {\includegraphics[width=\textwidth]{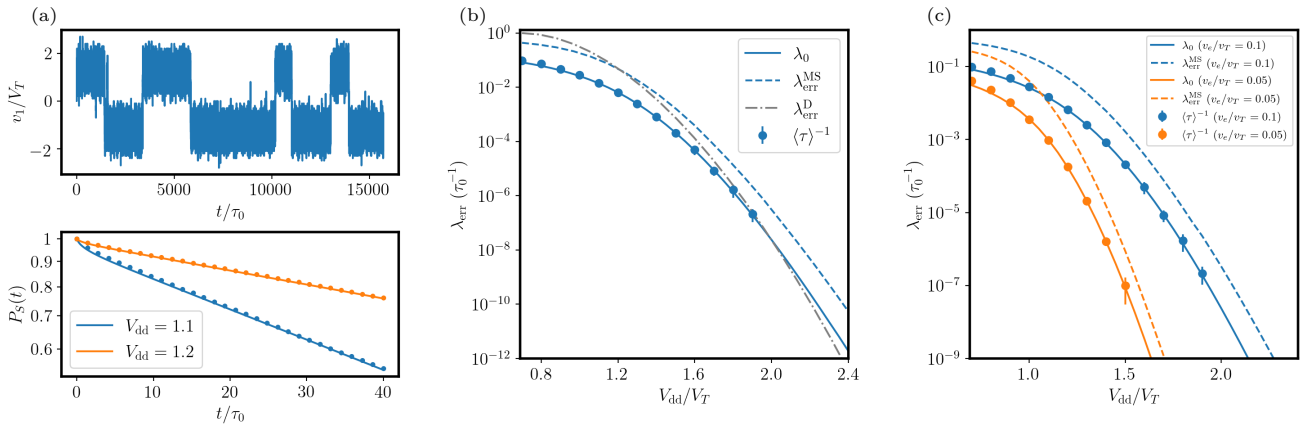}};
\node () at (-8.3,2.45) {\scriptsize (a)};
\node () at (-2.6,2.45) {\scriptsize (b)};
\node () at (3.6,2.45) {\scriptsize (c)};
\end{tikzpicture}
\caption{(a) Sample trajectory generated with the Gillespie simulation of the
stochastic dynamics (top, $V_\text{dd}/V_T=1.2$, $v_e/V_T=0.1$, $\text{n=1}$), and decay of the survival probability $P_S(t)$ for
the protocol described in the text, for different values of $V_\text{dd}$ (bottom).
Solid lines were obtained by Eq. \eqref{eq:survival} and the dots from data generated
with the Gillespie algorithm ($v_e/V_T=0.1$, $\text{n=1}$).
(b) Different estimates of the error rate as a function of $V_\text{dd}$ for $v_e/V_T=0.1$ and $\text{n=1}$. The dots indicate the inverse of the mean TTE, $\mean{\tau}^{-1}$,
as obtained from Gillespie simulations. The solid blue line corresponds to the minimum
eigenvalue $\lambda_0$ of the partial generator $-W_\text{HH}$, and the violet line to the metastable rate $\lambda_\text{err}^\text{MS}$ of Eq. \eqref{eq:lerr}. The dashed grey line shows the
dominant contribution in the $v_e /V_T \to 0$ limit of Eq. \eqref{eq:lerr_ld}.
(c) Estimates of the error rate as a function of $V_\text{dd}$ for different
values of $v_e/V_T$ ($\text{n=1}$).}
\label{fig:lerr}
\end{figure*}

The first step to compute $\lambda_\text{err}$ is to provide an operational definition
of what an error is. We consider that the state of the memory is read by monitoring
the output of the first inverter, i.e., the voltage $v_1$. A zero or positive value of $v_1$
is identified with the logical state $H$ (`high'), and a negative value with
the logical state $L$ (`low').
This logical encoding induces natural projection operations in the state space, that we construct
as follows. Each microscopic state $(v_1, v_2)$ is mapped to a vector $\ket{v_1,v_2}$.
A given probability distribution $P(v_1, v_2)$ is represented as
the vector $\ket{P} = \sum_{v_1, v_2} P(v_1, v_2) \ket{v_1, v_2}$, while the generator
of the master equation in Eq. \eqref{eq:master} is represented as a matrix $\mathds{W}$
acting over these vectors. Thus, the steady state distribution $\ket{P_\text{ss}}$ satisfies $0=\mathds{W}\ket{P_\text{ss}}$. The orthogonal projectors corresponding to the
logical states $H$ and $L$ are, respectively,
$\Pi_H = \sum_{v_1\geq0, v_2} \ket{v_1, v_2}\bra{v_1, v_2}$ and
$\Pi_L = \sum_{v_1<0, v_2} \ket{v_1, v_2}\bra{v_1, v_2}$ (where $\bra{a}$ is just the
transpose of $\ket{a}$).
Note that $\Pi_j \Pi_k = \delta_{j,k} \Pi_j$ and that $\Pi_H + \Pi_L = \mathds{1}$.
Then, we can consider the projections of the steady state to each of the logical
subspaces: $\ket{P_\text{ss}^H} = \Pi_H \ket{P_\text{ss}}/\bra{1}\Pi_H\ket{P_\text{ss}}$
and $\ket{P_\text{ss}^L} = \Pi_L \ket{P_\text{ss}}/\bra{1}\Pi_L\ket{P_\text{ss}}$ ($\ket{1}$
is just the vector with unit components). Now we give the following operational
definition of an error: at time $t=0$ we prepare the system at a state drawn
from the metastable distribution $\ket{P_\text{ss}^H}$ (for which the voltage $v_1$
is always positive or zero), and monitor its evolution until $v_1$ becomes negative.
This event is considered an error, and the random time $\tau$ at which
it takes place is recorded. We are interested in the distribution of $\tau$, which
can be considered a \emph{first-passage} problem \cite{van1992, redner2001}. As explained
it  \cite{van1992}, one possible approach to obtain the statistics of $\tau$ is to
consider an alternative dynamics with absorbing boundary conditions at the interface
between the logical subspaces. Thus, the \emph{survival probability} of not observing
any error up to time $t$ is given by
\begin{equation}
P_S(t) = \bra{1} e^{\mathds{W}_\text{HH} t} \ket{P_\text{ss}^H}.
\label{eq:survival}
\end{equation}
Here, the matrix $\mathds{W}_\text{HH}$ is the partial generator $\Pi_H \mathds{W} \Pi_H$ reduced
to the $H$-subspace. The vectors $\ket{1}$ and $\ket{P_\text{ss}^H}$ are also reduced to the same
subspace. The probability to observe an error between times $t$ and $t+dt$ is $p(t) dt$, where
$p(t) = -d_t P_S(t)$. Then the average time to an error (TTE) is
\begin{equation}
\mean{\tau} = \int_0^\infty \tau p(\tau) d\tau = \int_0^\infty P_S(\tau) d\tau .
\label{eq:mean_tte}
\end{equation}
At variance with the full generator $\mathds{W}$, the partial generator
$\mathds{W}_\text{HH}$ does not conserve probability (since it continuously leaks into the
$L$-subspace), and therefore its largest eigenvalue is strictly lower than $0$. Indeed,
we can write $P_S(t) = \sum_k C_k e^{-\lambda_k t}$, where $-\lambda_k$ are the
eigenvalues of $\mathds{W}_\text{HH}$ (with $ 0 < \lambda_0 \leq \lambda_1 \leq  \lambda_2 \leq \cdots$),
and $C_k$ are constants that depend on the initial state (with $\sum_k C_k=1$). Thus,
for large times we have $P_S(t) \simeq C_0 \exp(-\lambda_0 t)$. From this, it follows that for long times the distribution
of $\tau$ is approximately exponential with rate $\lambda_0$. This already provides a method
to estimate the error rate: one should construct the generator $\mathds{W}_\text{HH}$ and numerically
compute the eigenvalue of smallest absolute value, which can be done efficiently with several
routines since the matrix $\mathds{W}_\text{HH}$ is sparse. Note that $\lambda_0$ is
independent of the initial state. It is possible to obtain analytically another estimate of the
error rate by exploiting the metastability of the initial state $\ket{P_\text{ss}^H}$. For this,
we consider an approximation in which the state
$\ket{P(t)} = e^{\mathds{W}_\text{HH} t} \ket{P_\text{ss}^H}$ evolving according
to the generator $\mathds{W}_\text{HH}$ is assumed to be always proportional to
$\ket{P_\text{ss}^H}$ (the initial distribution), but with a time dependent normalization.
In that case the survival probability satisfies
$d_t P_S(t) = \bra{1} \mathds{W}_\text{HH} \ket{P_\text{ss}^H} P_S(t)$ and therefore
we can write $P_S(t) = e^{-\lambda_\text{err}^\text{MS}t}$, with the `metastable' rate
$\lambda_\text{err}^\text{MS} = - \bra{1} \mathds{W}_\text{HH} \ket{P_\text{ss}^H}$.
This is equivalent to assume that the error rate is constant and equal to the initial one, and
consequently depends explicitly on the initial state.
Note that by the conservation of probability of the full generator ($\bra{1}\mathds{W} = 0$), and
the property $\Pi_L + \Pi_H = \mathds{1}$,
we have the alternative expression $\lambda_\text{err}^\text{MS} =  \bra{1} \mathds{W}_\text{LH} \ket{P_\text{ss}^H}$, where $\mathds{W}_\text{LH}$ is the reduction of
the matrix $\Pi_L \mathds{W} \Pi_H$ to the appropriate subspaces.
This last expression for $\lambda_\text{err}^\text{MS}$ can be evaluated using Eq. \eqref{eq:ss_parity}
for the steady state, with the following result:
\begin{equation}
\lambda_\text{err}^\text{MS} = 4 \sum_{v_2} B(0, v_2) \: P(-v_2/2) \: Q(v_2/2),
\label{eq:lerr}
\end{equation}
where $B(v_1,v_2)$ is given in Eq. \eqref{eq:rates_up_down}, and $P(x)$ and $Q(y)$
are the LD approximations to the partial distributions, i.e., $P(x) \propto \exp(-g(x)/v_e)$
and  $Q(y) \propto \exp(-h(y)/v_e)$. It is instructive to see how Eq. \eqref{eq:lerr}
reduces to Eq. \eqref{eq:lerr_ld} for $v_e \to 0$. First, we notice that
$Q(y)$ becomes strongly peaked around $y=0$ for $v_e \to 0$, and therefore we
can approximate $\lambda_\text{err}^\text{MS} \simeq 4B(0,0)P(0)$. In turn, we have
$P(0) = \exp(-g(0)/v_e)/N$ with $N=\sum_{x} \exp(-g(x)/v_e)$, that for $v_e\to0$
becomes $N\simeq \exp(-g(x_\text{min})/v_e)$.
Then, we recover the result of Eq. \eqref{eq:lerr_ld},
with $\tau_0^{-1}$ replaced by the factor $4B(0, 0)$
(which is subexponential, since the rates scale as $v_e^{-1}$).

In general there is no definite relation between the estimates $\lambda_0$ and
$\lambda_\text{err}^\text{MS}$, and the mean TTE $\mean{\tau}$. However, for the particular protocol
we are considering, in which the initial state is $\ket{P_\text{ss}^H}$, the instantaneous
decay rate of the survival probability $\lambda(t) = -d_t \log(P_S(t))$
is a monotonously decreasing function. This is easily understood: the steady state distribution has a non-zero value at the boundary $v_1=0$ between logical subspaces. Then, the initial occupation of the states at or close to the boundary
will quickly leak into the $L$-subspace, with a rate that decreases as the occupation of those states decrease, reaching its asymptotic value $\lambda_0$
for long times.
In that case, from Eq. \eqref{eq:mean_tte} it follows that inverse of the
average TTE is bounded by $\lambda_0$ and $\lambda_\text{err}^{MS}$:
\begin{equation}
\lambda_0 \leq \mean{\tau}^{-1} \leq  \lambda_\text{err}^\text{MS}.
\end{equation}
Thus, $\lambda_\text{err}^\text{MS}$ provides an upper bound to the inverse mean TTE.

In Figure \ref{fig:lerr}-(a) we show a sample trajectory obtained by the Gillespie
algorithm, and the decay of the survival probability,
computed with two methods. The solid lines were obtained from Eq. \eqref{eq:survival}, by
constructing the reduced generator $W_\text{HH}$. The dots were obtained from Gillespie simulations
in which initial states were drawn from the steady state distribution and the time to an error
was recorded. We see that the decay rate decreases monotonously from the initial one
to the asymptotic one given by $\lambda_0$. From the same data we compute the mean TTE $\mean{\tau}$.
In Figure \ref{fig:lerr}-(b) we compare $\mean{\tau}^{-1}$ with the different estimates of the
error rate, as a function of $V_\text{dd}$.
We see that $\lambda_0$ is an excellent estimate of $\mean{\tau}^{-1}$.
The metastable rate $\lambda_\text{err}^\text{MS}$ of Eq. \eqref{eq:lerr} consistently overestimate the
true error rate, but displays the same scaling with $V_\text{dd}$. In contrast, we see that
the dominant estimate of Eq. \eqref{eq:lerr_ld} largely overestimate the error rate for low $V_\text{dd}$, while it underestimate it for large values of $V_\text{dd}$.
Figure \ref{fig:lerr}-(c) shows $\mean{\tau}^{-1}$, $\lambda_0$ and $\lambda_\text{err}^\text{MS}$
as a function of $V_\text{dd}$ for different values of $v_e$.

\begin{figure}[ht!]
\centering
\includegraphics[scale=.26]{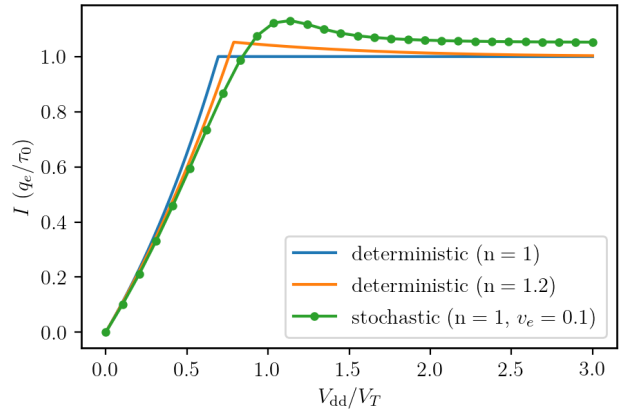}
\caption{Electrical current through each transistor in steady state conditions
as a function of $V_\text{dd}$.}
\label{fig:current}
\end{figure}

\section{Entropy production}

We now study the steady state entropy production
of the memory. At steady state, the average current $I$ through both inverters
is the same. Thus, the rate at which heat is dissipated in the environment is
$\dot Q = 4V_\text{dd} I$, and the entropy production rate is just
$\dot \Sigma = \dot Q/T$. From the deterministic solution for $\text{n}=1$,
it follows that in the monostable phase the electric current increases exponentially
with $V_\text{dd}$, $I=(q_e/\tau_0) (e^{V_\text{dd}/V_T}-1)$, while it is constant
in the bistable phase, $I=q_e/\tau_0$ (see Appendix \ref{ap:deterministic}). In Figure \ref{fig:current} we show
that the same constant value is achieved also for $\text{n} \neq 1$.
In addition, we show the average current obtained by computing the mean value of
$I(v_1,v_2) = q_e (\lambda_+^p(v_1,v_2)-\lambda_-^p(v_1,v_2))$ using the exact
steady state distribution $P_\text{ss}(v_1, v_2)$. This average current also
reaches a constant value for large $V_\text{dd}$, that is above the deterministic
one due to finite-$v_e$ effects. Interestingly, it displays a bump right after the onset
of bistability. The origin of this maximum in the average current is precisely
the occurrence of errors, since each switching event in which the memory flips
its state has an associated dissipation. As $V_\text{dd}$ increases, errors become
rare and the average current tends to the value corresponding to any of the metastable
NESSs with a definite logical value.

Thus, for large $V_\text{dd}$ the electrical current $I$ is just constant, and therefore
the entropy production $\dot \Sigma$ is proportional to $V_\text{dd}$.
Also, from Eq. \eqref{eq:exact_g} it is possible to see that, to dominant order in $V_\text{dd} \gg V_T$,
$\Delta g = g(0)-g(x_\text{min}) \simeq (2/(n+2)) (V_\text{dd}^2/V_T)$.
Then, it follows that for large entropy production
rates the error rate scales as:
\begin{equation}
\lambda^\text{MS}_\text{err} \propto e^{-\frac{2}{\text{n}+2}\: \frac{k_b T}{(4I)^2/C}
(\dot \Sigma/k_b)^2}.
\end{equation}
Here we have ignored terms in $\log(\lambda_\text{err})$ that are constant
or linear in $\dot \Sigma$, that can be easily included. Indeed, when expressed in terms of the voltage $V_\text{dd}$, the previous equation is compatible with what was obtained in ad-hoc treatments based on Gaussian noise \cite{natori1998}, up to model-dependent constant factors in the exponent. However, in general one must employ the result in Eq. \eqref{eq:lerr}, that can be readily evaluated.

\section{Discussion}

We used the theory of stochastic thermodynamics to construct a thermodynamically consistent stochastic model of a technologically relevant kind of electronic memory, subjected to Poissonian thermal noise. Utilizing the theory of large deviations, we obtained an analytical expression for the steady state of the memory which we used to estimate the rate at which errors occur. We have thus explicitly solved a problem that has been so far only treated using expensive numerical simulations \cite{rezaei2020}.

From a wider perspective, our work shows how modern developments in statistical physics can contribute to solve important problems in electronic engineering. Although our focus has been on the problem of memory reliability, our methods and results are also relevant for the design of non-conventional stochastic computing schemes, where naturally occurring thermal fluctuations are exploited as a resource \cite{palem2005, han2013, kaiser2020, borders2019, freitas2020}.
For instance, we note that our results directly apply to the low-power binary stochastic
neuron proposed in \cite{freitas2020} which is based on a SRAM memory cell core identical to the one studied here.

\section{Acknowledgments}

We acknowledge funding from the European Research Council, project NanoThermo (ERC-2015-CoGAgreement No. 681456), and from the Luxembourg National Research Fund (FNR), CORE project NTEC (C19/MS/13664907).

\bibliography{references.bib}

\onecolumngrid
\appendix

\section{Deterministic treatment of the CMOS SRAM cell}
\label{ap:deterministic}

In this section we derive the deterministic equations for a CMOS SRAM cell working
in the sub-threshold regime. We first consider a single inverter with input voltage
$v_g$ and output voltage $v$, and symmetric powering with voltages
$V_\text{dd} = -V_\text{ss}$. The current $I_p(v, v_g)$ through
the pMOS transistor for given $v$ and $v_g$ is \cite{enz2006}:
\begin{equation}
I_p(v,v_g) = I_0 e^{-V_\text{th}/V_T} \: e^{(V_\text{dd}-v_g)/(\text{n}V_T)} \: (1-e^{-(V_\text{dd}-v)/V_T}),\\
\end{equation}
while for the nMOS transistor we have $I_n(v, v_g) = I_p(-v, -v_g)$. From this we
can construct the deterministic dynamical equations for the voltages $v_1$ and $v_2$
of the CMOS SRAM cell discussed in the main text:
\begin{equation}
\begin{split}
C \: \frac{d v_1}{dt} &= I_p(v_1, v_2) - I_n(v_1,v_2)  \\
C \: \frac{d v_2}{dt} &= I_p(v_2, v_1) - I_n(v_2,v_1).
\end{split}
\label{apeq:determ_dyn}
\end{equation}
We first solve for the stationary solution satisfying $dv_1/dt = dv_2/dt = 0$.
By symmetry, this solution must satisfy $v_1 = -v_2 = v^*$. Thus, we need to find
$v^*$ such that $I_p(v^*, -v^*) = I_n(v^*, -v^*)$. In the following, for simplicity,
we consider the case $\text{n}=1$. In that case, the possible solutions are $v_0 = 0$ and, only if $V_\text{dd} > V_T \log(2)$,
\begin{equation}
v_{\pm} =  V_\text{dd} + V_T \log\left( 1/2 \pm \sqrt{1/4-e^{-2V_\text{dd}/V_T}}\right).
\end{equation}
Note that $v_+ = -v_-$, since actually these are the two solutions in the bistable phase.
We now consider $v_1 = v_+ + \delta v_1$ and $v_2 = v_- + \delta v_2$ and
expand Eq. \eqref{apeq:determ_dyn} to first order in $\delta v_{1/2}$, finding:
\begin{equation}
\frac{d}{dt}
\begin{bmatrix}
  \delta v_1\\
  \delta v_2
\end{bmatrix}
=
\frac{I_0e^{-V_\text{th}/V_T}}{CV_T}
\begin{bmatrix}
2-e^{2V_\text{dd}/V_T} & -2 \\
-2 & 2-e^{2V_\text{dd}/V_T}
\end{bmatrix}
\begin{bmatrix}
  \delta v_1\\
  \delta v_2
\end{bmatrix}
\end{equation}
The eigenvalues of the matrix in the previous equation are $-e^{2V_\text{dd}/V_T}$
and $4-e^{2V_\text{dd}/V_T}$, which shows that the solution considered is indeed
stable for $V_\text{dd} > V_T \log(2)$ (a similar analysis shows that the solution
$v_0$ becomes unstable at the same point), and that small departures relax back
to it at a rate
$\lambda_\text{eq} \simeq \tau_0^{-1} (v_e/V_T) \: e^{2V_\text{dd}/V_T}$,
with $\tau_0 = (q_e/I_0) e^{V_\text{th}/V_T}$.

From the previous solution it can be seen that the stationary current through each transistor
is $I_n = I_p = (q_e/\tau_0) (e^{V_\text{dd}/V_T}-1)$ for $V_\text{dd} \leq V_T \log(2)$
(monostability), and $I_n = I_p = q_e/\tau_0$ for $V_\text{dd} > V_T \log(2)$ (bistability). Thus, the current in the bistable phase is constant and the total entropy
production is $\dot \Sigma = 2 (2V_\text{dd}q_e/\tau_0)$.

\section{Macroscopic limit and large deviations principle}
\label{ap:large_dev}

The conduction channel of a MOS transistor in typical designs has two associated
dimensions: its width $W$ and its length $L$ \cite{tsividis2011, enz2006}.
The capacitance between the gate terminal and the body of the transitor (which is
typically the largest one) scales as the area of the channel: $C \propto W L$.
Also, the current through the channel for fixed drain-source and
gate-source voltages is proportional to the channel width, and inversely proportional
to the channel length \cite{enz2006}.
Thus, the parameter $I_0$ used to characterize the I-V curve
of the transistor scales as $I_0 \propto W/L$. For the following discussion we are
going to consider a family of devices with fixed channel length, but variable
channel width. Thus, we can consider $W$ as a scale parameter, with respect to which
both the capacitance $C$ and the current $I_0$ are proportional. In that case,
as considered in the main text, the elementary voltage $v_e = q_e/C$ scales as $W^{-1}$,
while the Poisson rates $\lambda_{\pm}^{n/p}(v_1, v_2)$ associated to the transistors scale as $W$. Under those conditions, the master equation in the main text can be rewritten as:
\begin{equation}
d_t P(\bm{v}, t) = \sum_\rho v_e^{-1} \left[
\omega_\rho(\bm{v} - v_e \bm{\Delta}_\rho, v_e)
P(\bm{v} - v_e \bm{\Delta}_\rho)
-\omega_\rho(\bm{v}, v_e) P(\bm{v}) \right].
\label{apeq:master_eq}
\end{equation}
In the previous equation, $\bm{v} = (v_1, v_2)^T$ is the state vector, and the
index runs over the possible transitions. For example, the values
$\rho=1,\cdots, 4$ correspond to the forward transitions of each transistor, while $\rho=-1,\cdots,-4$ to the reverse transitions.
The vectors $\bm{\Delta}_\rho$ encode the change in voltage associated to each transition.
The scaled rates $\omega_\rho(\bm{v}, v_e)$ are related to the original Poisson rates
$\omega_\rho(\bm{v}, v_e)$ by $\lambda_\rho(\bm{v}, v_e) = v_e^{-1}\omega_\rho(\bm{v}, v_e)$.
Thus, the scaling of the rates with respect to $W$ (or equivalently, with respect to $v_e$),
is taken into account in the factor $v_e^{-1}$, in such a way that the limit
$\lim_{v_e\to 0} \omega_\rho(\bm{v}, v_e)$ is well defined (the limit $v_e \to 0$ here and below must be interpreted as $v_e/V_T \to 0$
and $v_e/V_\text{dd} \to 0$ for fixed $V_T$ and $V_\text{dd}$).
Note that the explicit dependence
of the rates in the elementary voltage $v_e$ stems from the charging effects discussed
in the main text.

Under these conditions, the solution of the master equation in Eq. \eqref{apeq:master_eq}
satisfies a large deviations principle in the macroscopic limit $v_e \to 0$.
In order to see this, we introduce the large deviations ansatz
$P(\bm{v}, t) \asymp \exp(-(f(\bm{v},t)+o(v_e))/v_e)$ into Eq. \eqref{apeq:master_eq},
and only keep the dominant terms in $v_e \to 0$.
We note that in that limit
$P(\bm{v} - v_e \bm{\Delta}_\rho, t) \asymp P(\bm{v}, t)
\exp((\bm{\Delta}_\rho)_i \partial_{v_i} f(\bm{v},t))$. Therefore, the master
equation in Eq. \eqref{apeq:master_eq} reduces to the following dynamical equation
for the rate function:
\begin{equation}
d_t f(\bm{v}, t) = \sum_\rho \omega_\rho(\bm{v},0)
\left[1 - e^{(\bm{\Delta}_\rho)_i \partial_{v_i} f(\bm{v}, t)}\right].
\end{equation}
It is worth noting that for
general jump processes with scaling properties as the ones satisfied by
Eq. \eqref{apeq:master_eq}, the validity of the large deviation principle can be formally
proven \cite{cossetto2020}.

For the particular circuit under consideration, we can see from the previous
equation that the steady state rate function $f(v_1, v_2)$ should satisfy
\begin{equation}
    0 = \left(e^{\partial_{v_1} f }-1\right) \: a(v_1,v_2) + \left(e^{-\partial_{v_1} f}-1\right) \: b(v_1, v_2)  +
     \left(e^{\partial_{v_2} f}-1\right) \: a(v_2,v_1) + \left(e^{-\partial_{v_2}  f}-1\right) \: b(v_2, v_1),
\end{equation}
as presented in the main text, where the functions $a(v_1,v_2)$ and $b(v_1, v_2)$
were defined as the appropriate combination of the scaled transition rates.
The previous equation cannot be solved exactly. However, it can be employed to
solve for reduced rate functions derived from $f(v_1,v_2)$, exploiting the
symmetry of the problem and the contraction principle of large deviations theory.
We begin by changing variables to $x=(v_1 - v_2)/2$ and $y=(v_1+v_2)/2$.
Then, $\partial_{1/2} f = (\pm \partial_x f + \partial_y f)/2$. Defining
$\alpha = e^{\partial_x f/2}$ and $\beta = e^{\partial_y f/2}$, the previous equation
becomes:
\begin{equation}
    0 = \left(\alpha\beta-1\right) \: a(x,y) +
    \left(\alpha^{-1}\beta^{-1}-1\right) \: b(x,y) +
    \left(\beta/\alpha-1\right) \: a(-x,y) +
    \left(\alpha/\beta-1\right) \: b(-x,y),
    \label{apeq:new_var_rate_func}
\end{equation}
where the change of variables of the functions $a(x,y)$ and $b(x,y)$ is implicit.
Now, we are interested in computing the partial distributions $P(x)$ and
$Q(y)$ for the variables $x$ and $y$. The contraction principle states that
if the full distribution $P_\text{ss}(x,y)$ satisfies a large deviation principle
with rate function $f(x, y)$, then the partial distribution
$P(x) = \sum_{y} P_\text{ss}(x,y)$ also satisfies a large deviation principle
with rate function $g(x) = \inf_y f(x,y)$ \cite{touchette2009}. Then, assuming
that $f$ is sufficiently regular and that $\inf_y f(x,y) = \min_y f(x,y)$,
we have $g(x) = f(x, y_\text{min}|_x)$, where $y_\text{min}|_x$ is a minimum of $f(x,\cdot)$
and therefore satisfies $\partial_y f(x,y_\text{min}|x)=0$. Thus, $y_\text{min}|x$ is
the most probable value of $y$ for a given value of $x$. As discussed in the main
text, the variables $x$ and $y$ will always display some trivial, fine-grained correlations.
However, if those correlations are neglected, then $y_\text{min}|x$ becomes
independent of $x$ and, because of the symmetry of the system, it is actually equal to $0$. Thus, evaluating the previous equation at $y=0$, since $\beta|_{y=0} = 1$, we obtain:
\begin{equation}
\alpha|_{y=0} = e^{\partial_x f|_{y=0}/2} =  e^{d_x g(x)/2} =
\frac{a(-x,0) + b(x,0)}{a(x,0) + b(-x,0)},
\end{equation}
from where we easily obtain the expression for $d_x g(x)$ given in the main text. Note that from the previous expression is evident that
$d_x g(x)$ is an odd function, and therefore $g(x)$ is even. The
rate function $h(y)$ for the partial distribution $Q(y)$ can be obtained in a similar way. It is given by $h(y) =
f(x_\text{min}|y, y)$, where $x_\text{min}|y$ is a minimum of $f(\cdot, y)$. Again, neglecting correlations, we have that $x_\text{min}$ is independent of $y$, and can actually be computed
as the minimum of $g(x)$. In the bistable phase there are actually two equivalent values of $x_\text{min}$, that lead to the same function $h(y)$. Thus, evaluating Eq. \eqref{apeq:new_var_rate_func} at $x=x_\text{min}$, since $\alpha|_{x=x_\text{min}}=1$, we obtain:
\begin{equation}
\beta|_{x=x_\text{min}} =
e^{\partial_y f|_{x=x_\text{min}}/2} =  e^{d_y h(y)/2} =
\frac{b(x_\text{min},y) + b(-x_\text{min},y)}{a(x_\text{min},y) + a(-x_\text{min},y)}.
\end{equation}

For the Poisson rates corresponding to MOS transistors in subtreshold operation, that enter into the
definition of the functions $a(x,y)$ and $b(x,y)$, the integration of
$d_x g(x)$ can be performed exactly, leading to the compact expression given in the main text. This is not the case for $d_y h(y)$. However, it is possible to obtain the leading behaviour of $h(y)$ around $y=0$, which is given by:
\begin{equation}
h(y) =
\frac{2}{\text{n}}
\frac{(\text{n}-1)(1+e^{2(1+1/\text{n})x_\text{min}/V_T})+e^{(V_\text{dd}
+ x_\text{min})/V_T} + e^{(V_\text{dd} + x_\text{min}(1+2/\text{n}))/V_T}}
{1+e^{2(1+1/\text{n})x_\text{min}/V_T}+e^{(V_\text{dd} + x_\text{min})/V_T}
+ e^{(V_\text{dd} + x_\text{min}(1+2/\text{n}))/V_T}} \: y^2/V_T +\mathcal{O}(y^4).
\end{equation}

Finally, we note that the most probable values according to the large deviations
solution ($x=x_\text{min}$ and $y=0$, which correspond to $v_1 = -v_2 = x_\text{min}$)
match the deterministic solutions obtained in the previous section.

\end{document}